\documentclass[aps,prl,twocolumn]{revtex4}

\usepackage{amsmath}
\usepackage{graphicx}

\begin{document}

\title{Giant current-driven domain wall mobility in (Ga,Mn)As}

\author{Anh Kiet Nguyen, Hans Joakim Skadsem, and Arne Brataas}

\affiliation{Department of Physics, Norwegian University of Science
  and Technology, N-7491, Trondheim, Norway}

\begin{abstract}
  We study theoretically hole current-driven domain wall dynamics in
  (Ga,Mn)As. We show that the spin-orbit coupling causes significant
  hole reflection at the domain wall, even in the adiabatic limit when
  the wall is much thicker than the Fermi wavelength, resulting in
  spin accumulation and mistracking between current-carrying spins and
  the domain wall magnetization. This increases the out-of-plane
  non-adiabatic spin transfer torque and consequently the
  current-driven domain wall mobility by three to four orders of
  magnitude. Trends and magnitude of the calculated domain wall
  current mobilities agree with experimental findings.
\end{abstract}

\maketitle

Spin-polarized currents can reverse the magnetization, excite
spin-waves or move domain walls in ferromagnets. These are intriguing
phenomena which can become useful for magnetic memories. Ferromagnetic
semiconductors are especially interesting because the critical current
density ($j_c$) required to move domain walls is two to three orders
of magnitude smaller than in ferromagnetic
metals~\cite{Yamanouchi:Nature04,
  Yamanouchi:prl06,Gould:cm0602135,Yamaguchi:prl04}.  Much effort has
been invested experimentally~\cite{Freitas:jap85, Yamanouchi:Nature04,
  Yamanouchi:prl06, Gould:cm0602135, Yamaguchi:prl04, Chiba:prl06} and
theoretically~\cite{Berger:prb86,Slonczewski:jmmm96,Bazaliy:prb98,
  Tatara:prl04,Li:prl04,Zhang:prl04,Thiaville:epl05,Ohe:prl06,Xiao:prb06}
on current-driven domain wall dynamics~\cite{Marrows:aip05}. However,
it is not understood why the critical current density for domain wall
motion is so small in semiconductors compared to metals.

The spin-current induced torque on domain walls can be written as a
sum of an in-plane part and an out-of-plane part, $\boldsymbol{\tau
} \!\!=\!\!  \boldsymbol{\tau }^{in}\!+\!\boldsymbol{\tau }^{\perp}$. The
in-plane (out-of-plane) torque is spanned by (perpendicular to) the
gradient of the local magnetization. In ferromagnetic metals, the
domain wall width ($\lambda_w$) is large compared to the Fermi
wavelength ($\lambda_F$) and the spin-orbit coupling is weak. Here,
the domain wall does not reflect electrons which adiabatically align
their spin close to the local magnetization direction as they traverse
the domain wall. The associated angular momentum transfer induces an
in-plane torque on the domain wall~\cite
{Berger:prb86,Bazaliy:prb98,Tatara:prl04,Li:prl04,Zhang:prl04,
  Thiaville:epl05,Xiao:prb06}
\begin{equation}
  \boldsymbol{\tau }^{in}=-j_{s}\mathbf{m}\times \lbrack \mathbf{m}\times (
  \mathbf{\hat{\jmath}}\cdot \nabla )\mathbf{m}].  
\label{ATorque}
\end{equation}
$\boldsymbol{\tau }^{in}$ is also called the adiabatic spin torque.
In Eq.(\ref{ATorque}), $j_{s}$ is the spin current density. The unit
vectors $\mathbf{\hat{ \jmath}}$ and $\mathbf{m}$ are in the
directions of the current and the local magnetization, respectively.
At low current densities, $\boldsymbol{\mathcal{\tau }}^{in}$ does not
cause any steady-state motion of the domain
wall~\cite{Tatara:prl04,Li:prl04}.  However, even a small out-of-plane
torque $\boldsymbol{\tau }^{\perp}$ can induce a finite domain wall
drift velocity~\cite{Zhang:prl04, Thiaville:epl05, Xiao:prb06}. An
often used approximate form for the out-of-plane torque
is~\cite{Zhang:prl04, Thiaville:epl05}
\begin{equation}
  \boldsymbol{\tau }^{\perp}=-\beta j_{s}\mathbf{m} \times 
  (\mathbf{\hat{\jmath}}\cdot
  \nabla )\mathbf{m},  
\label{NATorque}
\end{equation}
where $\beta $ is a small dimensionless parameter.  The domain wall
drift velocity is controlled by the out-of-plane torque and is
proportional to $\beta $ \cite {Zhang:prl04,Thiaville:epl05}. In
literature, deviations from the adiabatic in-plane torque,
\textit{e.g.} $\boldsymbol{\tau }^{\perp}$, are often called the
non-adiabatic torque. We prefer to classify the torques as in-plane
and out-of-plane, where the latter can induce a steady state domain
wall motion.

In this Letter, we show that the intrinsic spin-orbit coupling in the
valence band of magnetic III-V semiconductors strongly enhances the
current-induced out-of-plane torque and thereby the domain wall
velocity. $\boldsymbol{\tau }^{\perp}$ depends non-locally on the
whole domain wall profile and cannot be described by the local
expression given by Eq.(\ref{NATorque}). We find that the steady-state
domain wall velocity ($v_{w}$) is proportional to the current density
($j$). The relevant measure of current-driven domain wall motion is
the domain wall current-mobility $\mu_{\scriptscriptstyle I} =
v_{w}/j$. Using realistic values for domain wall width and spin-orbit
coupling, we find that $\mu_{\scriptscriptstyle I}$ is enhanced by
three to four orders of magnitude compared to a system with weak or
vanishing spin-orbit coupling. This may explain an open question: why
$\mu_{\scriptscriptstyle I}$ (or $j_{c}$) is much larger (smaller) in
(Ga,Mn)As than in ferromagnetic metals~\cite{Yamanouchi:Nature04,
  Yamanouchi:prl06, Gould:cm0602135, Yamaguchi:prl04}.  

\begin{figure}[h]
\begin{picture}(100,85) 
   \put(-54,-13){\includegraphics[scale=0.27]{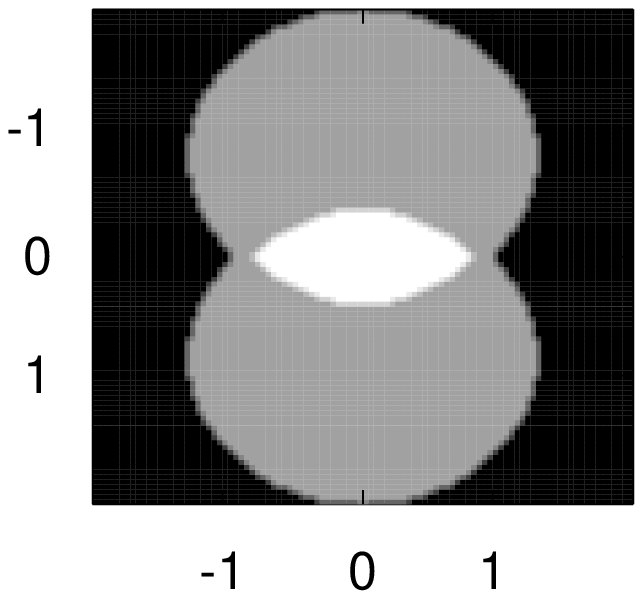}}
   \put(-9,-13){\includegraphics[scale=0.27]{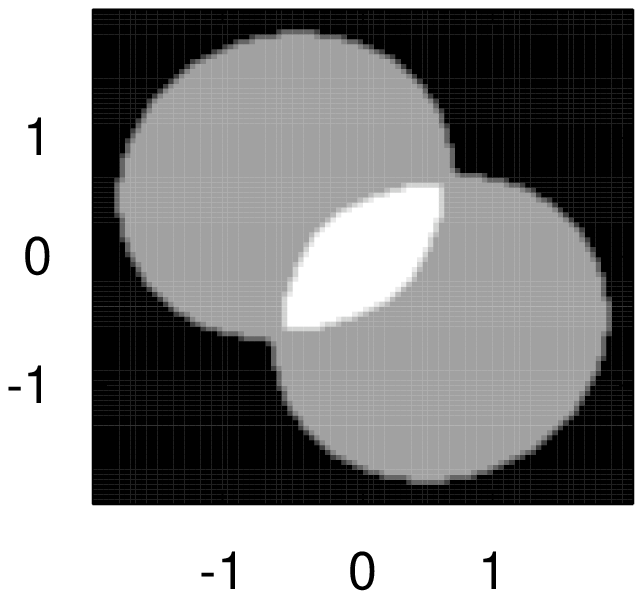}}
   \put( 36,-13){\includegraphics[scale=0.27]{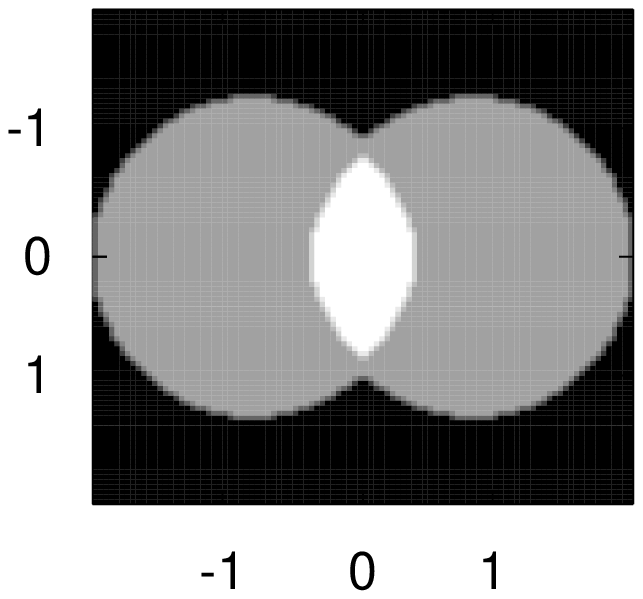}}
   \put( 81,-13){\includegraphics[scale=0.27]{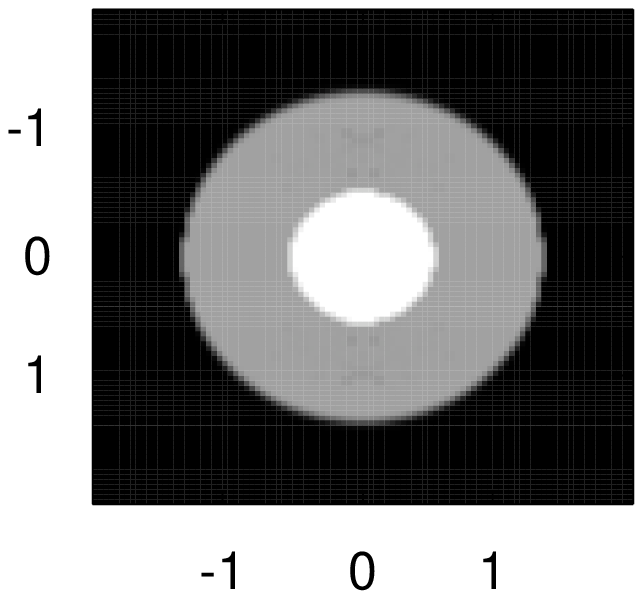}}
   \put(-25,50){a)}
   \put( 20,50){b)}
   \put( 65,50){c)}
   \put(120,50){d)}
   \put(-61,11){\rotatebox{90}{\small $k_z/k_{\scriptscriptstyle 0}$}}
   \put(-31,-10){\small $k_x/k_{\scriptscriptstyle 0}$}
   \put( 15,-10){\small $k_x/k_{\scriptscriptstyle 0}$}
   \put( 60,-10){\small $k_x/k_{\scriptscriptstyle 0}$}
   \put(107,-10){\small $k_x/k_{\scriptscriptstyle 0}$}
   \qbezier[30](-60,65)(-50,65)(-40,65)
   \qbezier[30](-60,65)(-60,75)(-60,85)
   \put(-60,85){\vector(0,1){2}}
   \put(-40,65){\vector(1,0){2}}
   \put(-65,83){$z$}
   \put(-45,59){$x$}
   \put(-20,65){\thicklines \vector(0,1){20}}
   \put( 25,65){\thicklines \vector(-1,1){14.15}}
   \put( 70,65){\thicklines \vector(-1,0){20}}
   \put(115,65){\thicklines \vector( 0,1){20}}
   \put(115,65){\thicklines \vector(-1,2){8.9}}
   \put(115,65){\thicklines \vector(-1,1){14.15}}
   \put(115,65){\thicklines \vector(-2,1){17}}
   \put(115,65){\thicklines \vector(-1,0){20}}
   \put(115,65){\thicklines \vector(-2,-1){17}}
   \put(115,65){\thicklines \vector(-1,-1){14.15}}
   \put(115,65){\thicklines \vector(-1,-2){8.9}}
   \put(115,65){\thicklines \vector( 0,-1){20}}
   \put(115,65){\oval(40,40)[bl]}
   \put(115,65){\oval(40,40)[tl]}
   \put(114,85.5){\vector(-4,-1){6}}
   \put(106,47.0){\vector(4,-1){6}}
\end{picture}
\caption{a), b) and c) Distributions of transverse propagating modes
  in (Ga,Mn)As for different magnetization directions shown by the
  arrows.  White (gray) color indicates that two (one) spin channels
  are open for transport.  d) Distribution of transverse conducting
  channels through an adiabatic Bloch domain wall in (Ga,Mn)As. Arrows
  show the magnetization directions itinerant holes encounter while
  traversing a Bloch wall. For an adiabatic domain wall, transverse
  modes in a) that do not exist in d) are total reflected by the wall.
  Such reflections increase the out-of-plane torque on the domain wall
  which subsequently increases the domain wall drift velocity. The
  plots are created using Eq.(\ref{Hamiltonian}) with $\gamma_1  \!\!=\!\!
   6.8$, $\gamma_2 \!\!=\!\! 2.7$,
  $h_{ex}/\epsilon_{\scriptscriptstyle 0} \!\!=\!\! 1.5$, $\epsilon_{
    \scriptscriptstyle F} / \epsilon_{\scriptscriptstyle 0} \!\!=\!\!
  2.25$.}
\label{PropModes}
\end{figure}

In conventional ferromagnetic metals only a very small fraction of the
electrons contribute to the out-of-plane torque via spin-flip
scattering at extrinsic impurities or via non-adiabatic corrections
due to a finite domain wall width~\cite{Zhang:prl04,
  Thiaville:epl05,Xiao:prb06}.  Radically different from this, the
strong intrinsic hole spin-orbit coupling in (Ga,Mn)As causes a finite
domain wall resistance even in the adiabatic limit ($\lambda_w \gg
\lambda_F$), by preventing a large fraction of holes to adiabatically
adjust their spins to the magnetization of the domain
wall~\cite{Nguyen:prl06a,Oszwaldowski:cm0605230}.  The intrinsic
spin-orbit coupling in combination with the magnetization cause an
anisotropic momentum distribution of propagating modes.
Fig.\ref{PropModes} a), b) and c) show distributions of transverse
propagating modes in (Ga,Mn)As for different magnetization
directions~\cite{Nguyen:prl06a}.  Only transverse modes that exist for
all magnetization directions within the domain wall,
Fig.\ref{PropModes} d), will conduct. The remaining modes are totally
reflected~\cite{Nguyen:prl06a,Oszwaldowski:cm0605230}.  The fraction
of reflected holes is large and comparable to the total number of
conducting particles. This induces an enhanced domain wall resistance,
as well as spin accumulation and mistracking between hole spins and
the magnetization of the domain wall. We find that the total
out-of-plane torque increases by three to four orders of magnitude
compared to systems with vanishing spin-orbit coupling. The
current-induced mobility is subsequently enhanced approximately
proportional to the total out-of-plane torque.

We shall use the simplest model that captures the essential physics of
holes in (Ga,Mn)As, namely the 4-band Kohn-Luttinger Hamiltonian in the
spherical approximation~\cite{Luttinger:pr56, Schliemann:cm0604585,
  Nguyen:prl06a, Jungwirth:rmp06}
\begin{equation}
  H=\frac{\hbar^{2}}{2m}\left[ \left( \gamma_{1}+\frac{5}{2}\gamma_{2}\right) 
    p^{2}-2\gamma_{2}(\mathbf{p}\cdot \mathbf{S})^{2}\right] -
  \mathbf{h}(\mathbf{r})\cdot \mathbf{S},  
\label{Hamiltonian}
\end{equation}
where, $m$ and $\mathbf{p}$ are the bare electron mass and the hole
momentum operator, respectively. $\mathbf{S}$ is a vector of $4\times
4$ dimensionless angular momentum operators for a $S \!\!=\!\! 3/2$ spin, and
$ \gamma_{1}$ and $\gamma_{2}$ are Luttinger parameters. In
Eq.\eqref{Hamiltonian}, $\mathbf{h}$ represents the exchange field
from the localized magnetic moments. The exchange field is assumed to
have a constant modulus and to be homogeneous in the transverse
directions, $|\mathbf{h}(y)| \!\!=\!\! h_{ex}$. The parameter $\gamma_{2}$
determines the strength of the effective spin-orbit coupling of the
holes. For a given doping density, Eq.\eqref{Hamiltonian} with
$\mathbf{h} \!\!=\!\! 0$ describes hole doped GaAs and provides a Fermi energy
$\epsilon_{\scriptscriptstyle 0}$, a Fermi wavevector
$k_{\scriptscriptstyle 0} \!\!=\!\! \sqrt{2 m_{\scriptscriptstyle 0}
  \epsilon_{\scriptscriptstyle 0}} / \hbar$ and a Fermi wave length
$\lambda_{\scriptscriptstyle 0} \!\!=\!\! 2\pi / k_{\scriptscriptstyle 0}$ for
heavy holes. Here, $m_{\scriptscriptstyle 0} \!\!=\!\! m/2.6$ represents the
heavy hole mass in GaAs.

Focusing on low temperatures and low bias transport properties, we
consider the \emph{linear response} at a Fermi energy
$\epsilon_{\scriptscriptstyle F}$. Our system is a discrete,
rectangular conductor of lengths $L_{x}$, $L_{y}$, $L_{z}$ and lattice
constants $a_{x}$,$a_{y}$,$a_{z}$ sandwiched between two reservoirs.
Born-von Karman boundary condi-tions are assumed in the transverse
directions $x$ and $z$.

We model the dynamics of domain walls by the dimensionless
Landau-Lifshitz-Gilbert (LLG) equation
\begin{equation}
  \frac{d\mathbf{m}}{d\tilde{t}}=-\mathbf{m} \times 
  \tilde{\mathbf{h}}^{\mathit{eff}}+\alpha \mathbf{m} \times 
  \frac{d\mathbf{m}}{d\tilde{t}} \, ,   
\label{LLG}
\end{equation}
where $\mathbf{m}(y)$ should be understood as a unit macrospin for the
transverse slice located at $y$, and $\alpha $ is the Gilbert damping
constant. The dimensionless time is $\tilde{t} \!\!=\!\!
t/t_{\scriptscriptstyle 0}$ where $t_{\scriptscriptstyle 0} \!\!=\!\!
M_{s}(\lambda_{\scriptscriptstyle 0}/2\pi )^{2}/2\gamma A$. Here,
$M_{s}$, $A$ and $\gamma \!\!=\!\! e/mc~$ are the saturation
magnetization, spin stiffness and the gyromagnetic ratio,
respectively. The effective magnetic field reads
\begin{equation}
  \tilde{\mathbf{h}}^{\mathit{eff}}=\tilde{\nabla}^{2}\mathbf{m}+\tilde{
    \mathbf{h}}^{an}\cdot \mathbf{m}-h_{pd}\boldsymbol{\chi }_{S}\,,
\label{Heff}
\end{equation}
where $\tilde{\nabla} \!\!=\!\! (\lambda_{\scriptscriptstyle 0}/2\pi
)\nabla $. $\tilde{\mathbf{h}}^{\mathit{an}}$ is the dimensionless
anisotropy field, which is used to control the type and the width of
the domain wall.  Demagnetization fields for simple geometries can be
included in $\tilde{ \mathbf{h}}^{\mathit{an}}$. The last term in
Eq.\eqref{Heff} is the contribution to the spin-transfer torque from
the non-equilibrium itinerant holes \cite{Nunez:cm0403710,Ohe:prl06}.
The dimensionless coupling constant $h_{pd}$ and the spin-density
response function $\boldsymbol{\chi }_{S}$ are defined below. The LLG
equation is numerically integrated using a fourth-order Runge-Kutta
method~\cite{Sun:prb00}.

We calculate the current carrying wave function by a stable transfer
matrix formalism~\cite{Usuki:prb95,Nguyen:prl06a}. The non-equilibrium
spin density $\langle \rho \mathbf{S}\rangle ^{\mathit{ne }}$ is
determined as a trace over all hole states between $\epsilon_{
  \scriptscriptstyle F}$ and $\epsilon_{\scriptscriptstyle F}\!+\!eV$ in
the left reservoir. We define the linear response function for the
spin density as $\boldsymbol{\chi}_{S}\!\!=\!\!(e \hbar
k_{\scriptscriptstyle 0}/j m_{\scriptscriptstyle 0})\langle \rho
\mathbf{S}\rangle ^{\mathit{ne}}$, which gives $h_{pd}\!\!=\!\!\hbar
h_{ex}\lambda_{\scriptscriptstyle 0}j/8\pi
eA\epsilon_{\scriptscriptstyle 0}$, that is proportional to the
current density $j$.
Similarly, we also compute the non-equilibrium spin current density,
$\mathbf{j}_{s}$, and the average spin per conducting holes, $\langle
\mathbf{S}(y)\rangle \!\!=\!\!\mathbf{j}_{s}e/j\hbar $.

The parameters describing our conductor are $a_x/\lambda_{
  \scriptscriptstyle 0} \!\!=\!\! a_z/\lambda_{ \scriptscriptstyle 0}
\!\!=\!\! 0.2/2\pi$, $a_y/\lambda_{ \scriptscriptstyle 0} \!\!=\!\!
0.75/2\pi$, $L_x \!\!=\!\! L_z \!\!=\!\! 51 a_x$ and $L_y \!\!=\!\!
400 a_y$. Furthermore, $h_{ex}/\epsilon_{\scriptscriptstyle 0}
\!\!=\!\!  1.5$ and $\epsilon_{\scriptscriptstyle F} /
\epsilon_{\scriptscriptstyle 0} \!\!=\!\! 2.25$. The ratio
$\epsilon_{\scriptscriptstyle F} / h_{ex} \!\!=\!\!  1.5$ is in the
same order as experimental values~\cite{Jungwirth:rmp06}.  The
anisotropy/demagnetization field $\tilde{h}^{\mathit{an}}_x \!\!=\!\!
0$, $\tilde{h}^{\mathit{an}}_y \!\!=\!\!  -1$ for simplicity and to
ensure that the considered domain walls are Bloch walls, consistent
with recent experiments~\cite
{Yamanouchi:Nature04,Yamanouchi:prl06,Chiba:prl06}. We use
$\tilde{h}^{\mathit{an}}_z$ to control the domain wall width
$\lambda_w/\lambda_{\scriptscriptstyle 0} \!\!=\!\!
\sqrt{1/\tilde{h}^{\mathit{an}}_z} / 2\pi$. The Gilbert damping
constant is assumed to be $\alpha \!\!=\!\!
0.03$~\cite{Sinova:prb04}.

The transfer of momentum and angular momentum from holes to the domain
wall must be treated on an equal footing. We achieve this as follows:
First, we find the equilibrium ($j \!\!=\!\! 0$) domain wall
configuration by integrating Eq.\eqref{LLG} with $ h_{pd}\!\!=\!\!0$
to a sufficiently large time. The resulting equilibrium domain wall
configuration is $m_{x}(y) \!\!=\!\! 1/\cosh (y/\lambda_{w})$,
$m_{y}(y) \!\!=\!\! 0$ and $m_{z}(y) \!\!=\!\! \tanh (y/\lambda_{w})$,
as expected from an analytical equilibrium solution of Eq.\eqref{LLG}.
Next, a constant current is applied by choosing a finite $h_{pd}$, and
the following two steps are iterated: 1) compute $\boldsymbol{\chi
}_{S}$ for a given $\mathbf{h} \!\!=\!\! h_{ex}\mathbf{m}$ using
Eq.\eqref{Hamiltonian}, and 2) integrate $\mathbf{m} \!\!=\!\!
\mathbf{h}/h_{ex}$ a time $\Delta \tilde{t}$ using Eq.\eqref{LLG}. We
use $\Delta \tilde{t} \!\!=\!\! 0.2$, which is sufficient for
convergence.

%
\begin{figure}[h]
\begin{picture}(100,210) 
    \put(-75,120){\includegraphics[bb=50 110
      410 302, clip=true,scale=0.5]{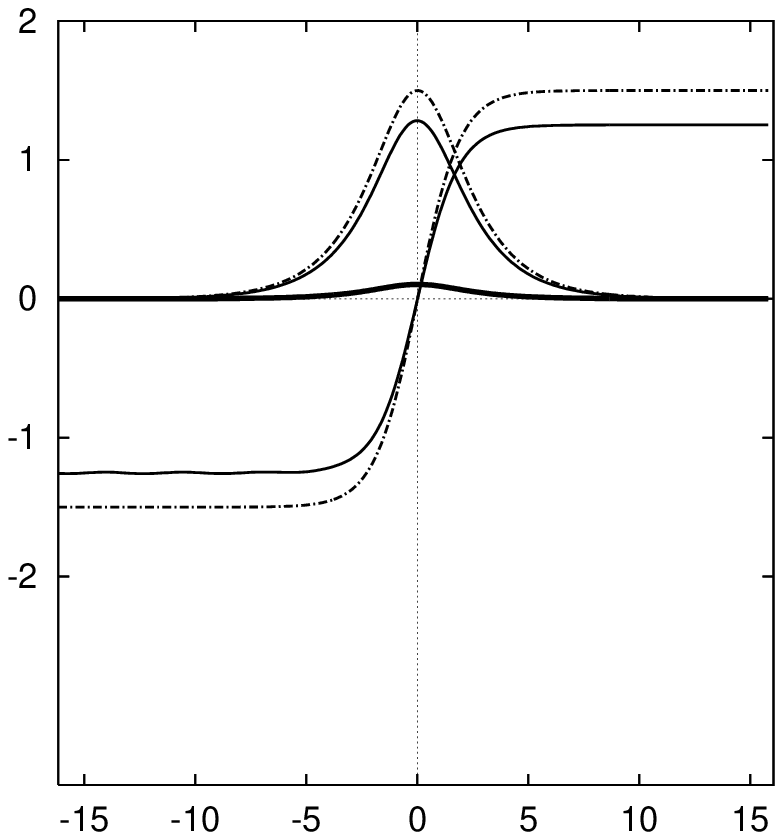}}
    \put(-75,0){\includegraphics[trim=0 0 0
      1,clip=true,scale=0.5]{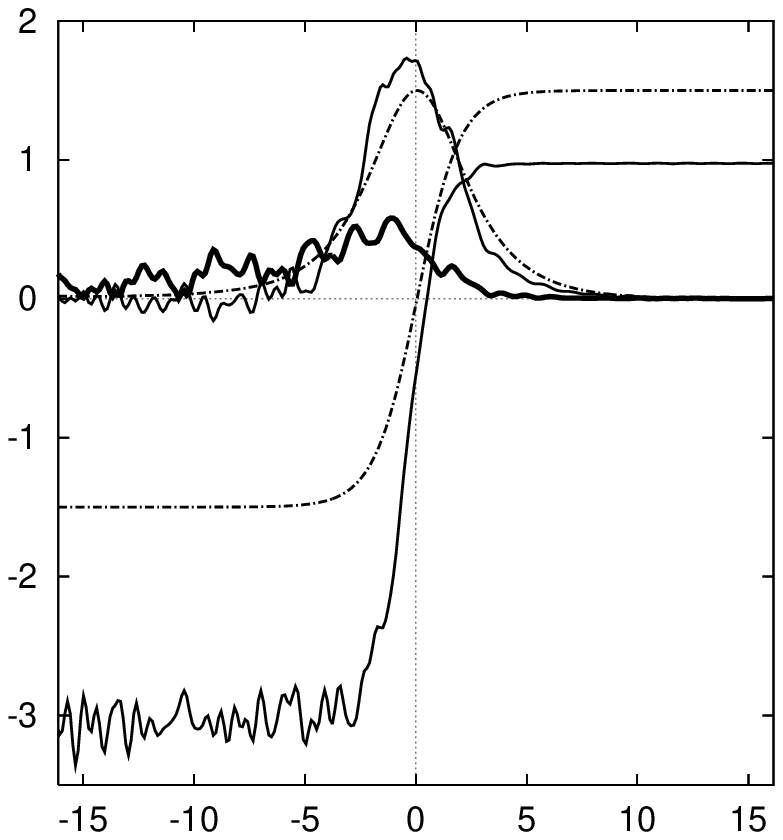}}
    \put(65,0){\includegraphics[scale=0.6]{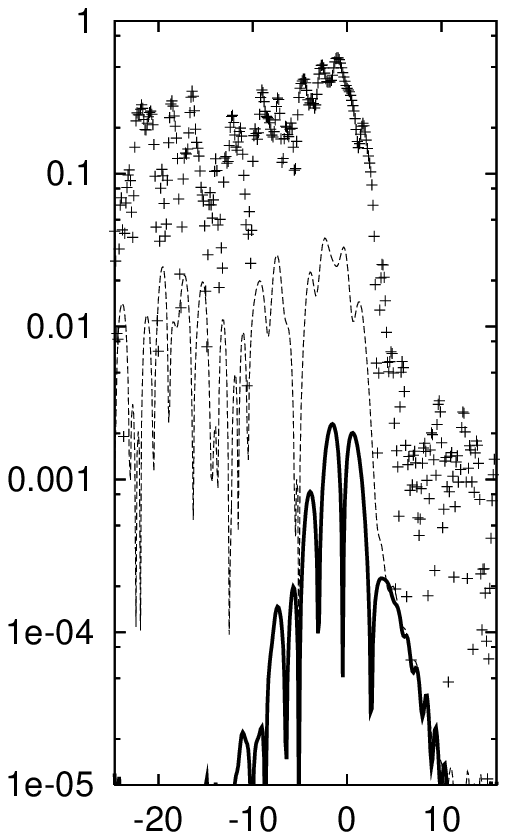}}
   \put(15,140){a)}
   \put(15,130){$\gamma_2 \!\!=\!\! 0$}
   \put(30,202){$h_z$}
   \put(15,180){$h_x$}
   \put(13,182){\vector(-1,0){10}}
   \put(-35,200){\footnotesize $\langle S_x \rangle$}
   \put(-25,196){\vector(2,-1){12}}
   \put(-55,148){\footnotesize $\langle S_z \rangle$}
   \put(0,154){\scriptsize $|\boldsymbol{\mathcal{T}}|$}
   \put(5,160){\vector(-1,2){6}}
   \put(15,30){b)} 
   \put(15,20){$\gamma_2 \!\!=\!\! 2$} 
   \put(30,112){$h_z$}
   \put(19,87){$h_x$} 
   \put(17,89){\vector(-1,0){10}}
   \put(-33,110){\footnotesize $\langle S_x \rangle$}
   \put(-55,25){\footnotesize $\langle S_z \rangle$}
   \put(0,63){\scriptsize $|\boldsymbol{\mathcal{T}}|$}
   \put(5,70){\vector(-1,2){6}}
   \put(-15,-5){$y/\lambda_{\scriptscriptstyle 0}$}
   \put(148,130){c)}
   \put(120,-5){$y/\lambda_{\scriptscriptstyle 0}$}
   \put(70,73){\rotatebox{90}{$| \mathcal{T}_y |$}}
\end{picture}
\caption{Steady-state snapshots. a) and b): exchange field
  ($h_{x}/\protect \epsilon_{\scriptscriptstyle 0}$ and
  $h_{z}/\protect \epsilon_{ \scriptscriptstyle 0}$), spin per
  conducting hole ($\langle S_{x}\rangle $ and $\langle S_{z}\rangle
  $) and absolute values of spin-torque per conducting hole $|
  \boldsymbol{\mathcal{T}}|$ as functions of position, for a)
  vanishing and b) finite spin-orbit coupling. c): absolute values of
  the out-of-plane torque $|\mathcal{T}_y|$ versus position for
  $\protect \gamma_{2}\!\!=\!\!0$ (thick solid line), $\protect
  \gamma_{2}\!\!=\!\!0.5$ (dashed line) and $\protect \gamma
  \!\!=\!\!2$ (symbol +).  For all plots, $h_{pd}\!\!=\!\!0.001$ and
  $\protect \lambda_{w}/ \protect \lambda_{\scriptscriptstyle
    0}\!\!=\!\!2$. }
\label{Torque}
\end{figure}
%
Let us first consider a system without spin-orbit coupling,
$\gamma_2\!\!=\!\!0$. We consider a
$\lambda_{w}/\lambda_{\scriptscriptstyle 0}\!\!=\!\!2$ domain wall,
close to the experimental values
\cite{Yamanouchi:Nature04,Yamanouchi:prl06}. Numerical calculation of
the conductance shows that only $0.1\%$ of the incoming holes are
reflected back by the domain wall.  From Fig.\ref{Torque}a), we see
that the hole spins follow the domain wall magnetization closely, and
$\langle \mathbf{S} (y)\rangle $ is virtually parallel to
$\mathbf{h}(y)$ throughout the system.  Outside the wall, $\langle
S_{z}\rangle \!\approx\! \pm 1.25$, since both the heavy ($S_{z}
\!\approx\!   -1.5$) and light ($S_{z} \!\approx\!  -0.5$) holes
participate in the transport. The absolute values of the dimensionless
spin-torque per conducting holes,
$\boldsymbol{\mathcal{T}}\!\!=\!\!(\mathbf{h}
/\epsilon_{\scriptscriptstyle 0})\times \boldsymbol{\chi }_{S}$, is
shown in Fig.\ref{Torque}a). It agrees well with the in-plane torque
given by Eq.(\ref{ATorque}) through the relation
$\boldsymbol{\mathcal{T}} \!\!=\!\!\boldsymbol{\tau }^{in}e\lambda_{
  \scriptscriptstyle 0}/2\pi j\hbar $. The out-of-plane torque is very
small, as expected for this case.

As shown in Fig.\ref{Torque}b), turning on the spin-orbit coupling
completely changes the physical picture. Numerical calculation of the
conductance shows that the domain wall reflects $45\%$ of the holes,
causing spin accumulation and mistracking between carrier spins and
the magnetization of the domain wall, particularly on the upstream
side of the wall. Interference between incoming and reflected holes
creates a spin-wave pattern in $\langle \mathbf{S}\rangle $ and
$\boldsymbol{\mathcal{T}}$, causing the reproducible ``noise'' in the
figures. The shape of the domain wall is, however, virtually
unchanged, due to the small current applied, $h_{pd}\!\!=\!\!0.001$. From our
previous study, we know that the majority of reflected holes consists
of heavy holes~\cite{Nguyen:prl06a}, in agreement with $\langle
S_{z}\rangle \!\approx\! -3.0$ on the left side of the domain wall.

As discussed in the introduction, the steady state domain wall
velocity is controlled by the out-of-plane torque \cite
{Zhang:prl04,Thiaville:epl05}. Fig.\ref{Torque}c) shows the absolute
value of the out-of-plane torque component, $|\mathcal{T}_{y}|$, for
three different spin-orbit couplings. The spatial total of
$|\mathcal{T} _{y}|$ denoted as $\langle |\mathcal{T}_{y}|\rangle
\!\!=\!\!  \int dy |\mathcal{T}_y|~$ may be used as an estimate of the
out-of-plane torque contributions to the domain wall drift velocity.
Note that even when the spin-orbit coupling vanishes, the out-of-plane
torque is finite since the system is not entirely adiabatic, $\lambda
_{w}/\lambda_{\scriptscriptstyle 0}\!\!=\!\!2\ll \infty$. Here, we
find that $|\mathcal{T}_{y}| \! \approx \! 1\%$ of the total torque
$|\mathcal{T}|$.  When the spin-orbit coupling $\gamma_{2}$ increases
from 0 to 2.7, the spatial total out-of-plane torque $\langle
|\mathcal{T}_{y}|\rangle $ increases by a factor of 1000.  This
explains the dramatic increase in the domain wall drift velocity and
thereby mobility.

\begin{figure}[h]
  \includegraphics[scale=0.5]{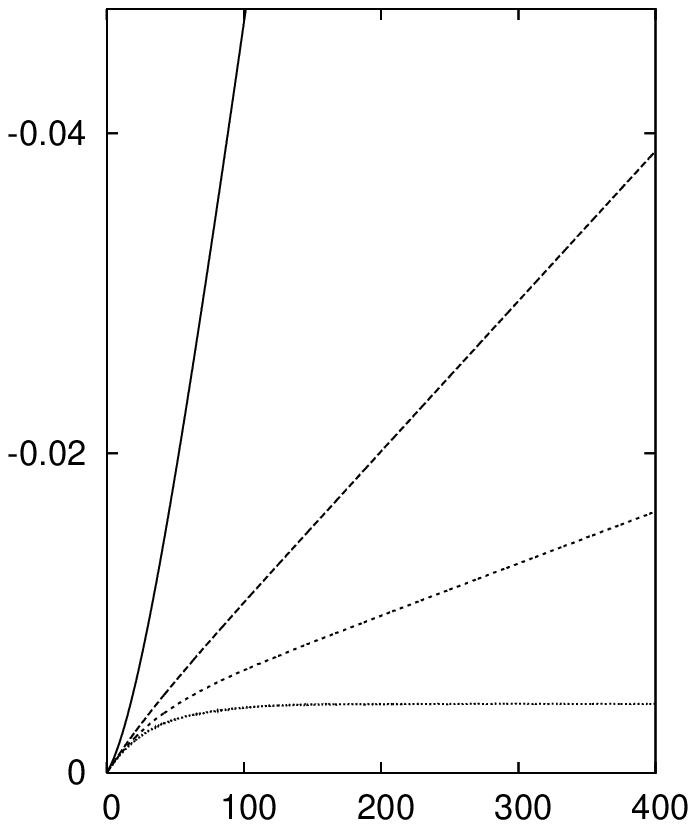}
  \includegraphics[scale=0.5]{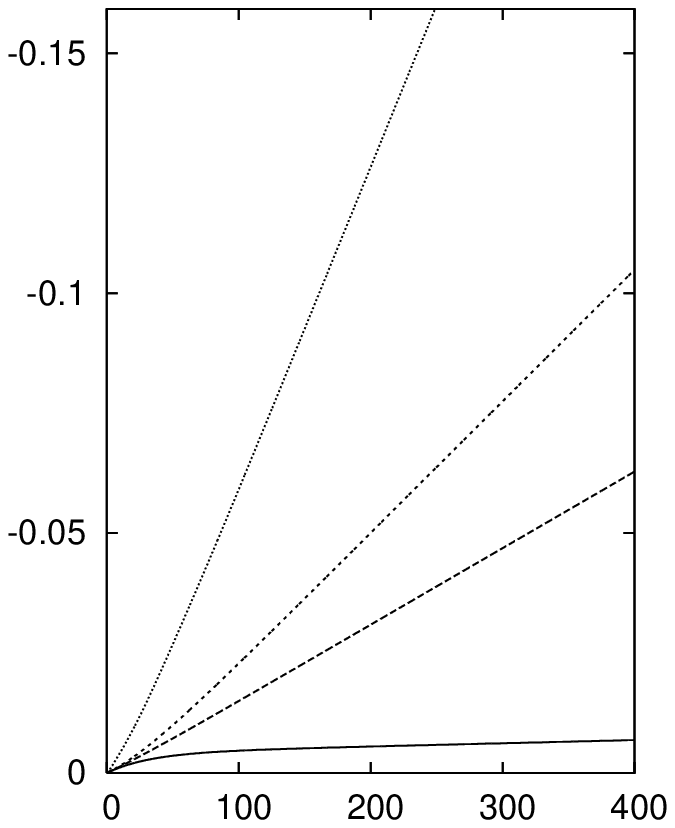} \put(-170,110){a)}
  \put(-170,103){$\gamma_2\!\!=\!\!0$} \put(-225,50){\rotatebox{90}{ $x_w /
      \lambda_{\scriptscriptstyle 0}$}} \put(-165,-5){$t /
    t_{\scriptscriptstyle 0}$}
  \put(-192,80){\rotatebox{77}{\scriptsize $0.5$}}
  \put(-168,55){\rotatebox{47 }{\scriptsize $\lambda_w /
      \lambda_{\scriptscriptstyle 0} \!\!=\!\! 1.0$}}
  \put(-140,42){\rotatebox{25}{ \scriptsize $1.5 $}}
  \put(-140,22){\rotatebox{0}{\scriptsize $3.0$}} \put(-75,110){b)}
  \put(-75,103){$\lambda_w / \lambda_{\scriptscriptstyle 0} \!\!=\!\! 2$}
  \put(-109,50){ \rotatebox{90}{$x_w / \lambda_{\scriptscriptstyle
        0}$}} \put(-50,-5){$t / t_{\scriptscriptstyle 0}$}
  \put(-70,60){ \rotatebox{65}{\scriptsize $0.35$}}
  \put(-60,40){\rotatebox{45}{\scriptsize $ \gamma_2\!\!=\!\!0.25$}}
  \put(-25,48){\rotatebox{25}{\scriptsize $0.20$}}
  \put(-25,16){\rotatebox{0}{\scriptsize $0.00$}}
  \caption{Domain wall displacement as a function of time for a) zero
    spin-orbit coupling for different domain wall widths, and b) fixed
    domain wall width for varying hole spin-orbit coupling strengths.
    For both plots, $ h_{pd}\!\!=\!\!0.001$.}
\label{DWPosition}
\end{figure}
%
Fig.\ref{DWPosition}a) displays the domain wall displacement ($x_w$)
as functions of time for varying domain wall widths for vanishing
spin-orbit coupling. $x_w$ completely saturates in the adiabatic
limit, $\lambda_w / \lambda_{\scriptscriptstyle 0} > 3$, as
expected~\cite{Tatara:prl04,Li:prl04,Ohe:prl06}. Beyond the adiabatic
limit, an increasing fraction of holes will be reflected, causing a
out-of-plane torque on the domain wall and a finite drift velocity,
which, as can be seen in Fig.\ref{DWPosition}a), increases for
decreasing domain wall width.

Let us now consider a finite spin-orbit coupling.
Fig.\ref{DWPosition}b) shows the domain wall displacement as a
function of time for different spin-orbit couplings. We see that a
larger spin-orbit coupling also increases the domain wall drift
velocity due to the increased reflection of holes. The reflected holes
increase the spin accumulation and mistracking, which, in turn,
increases the out-of-plane spin-torque on the domain wall and thereby
its drift velocity.

\begin{figure}[h]
  \includegraphics[scale=0.42]{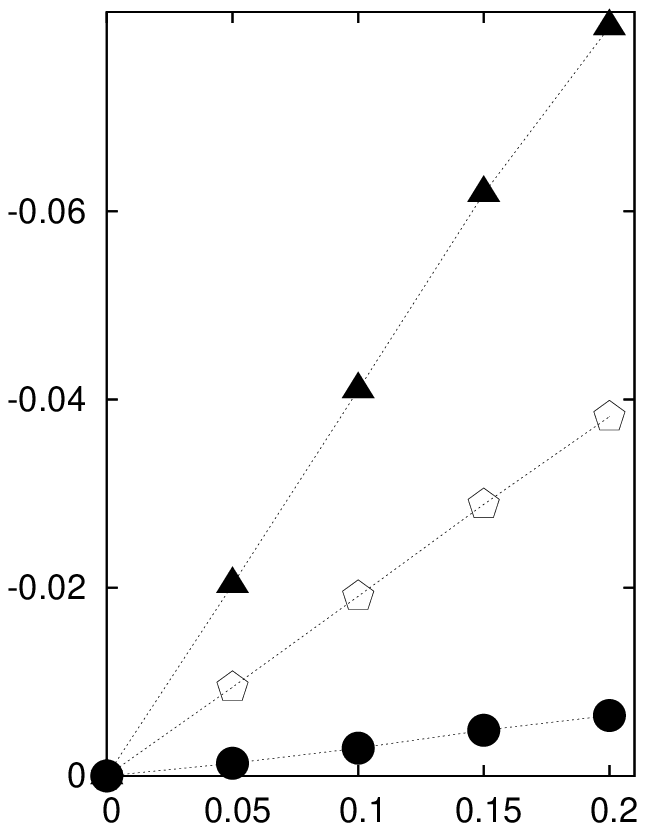}
  \includegraphics[scale=0.42]{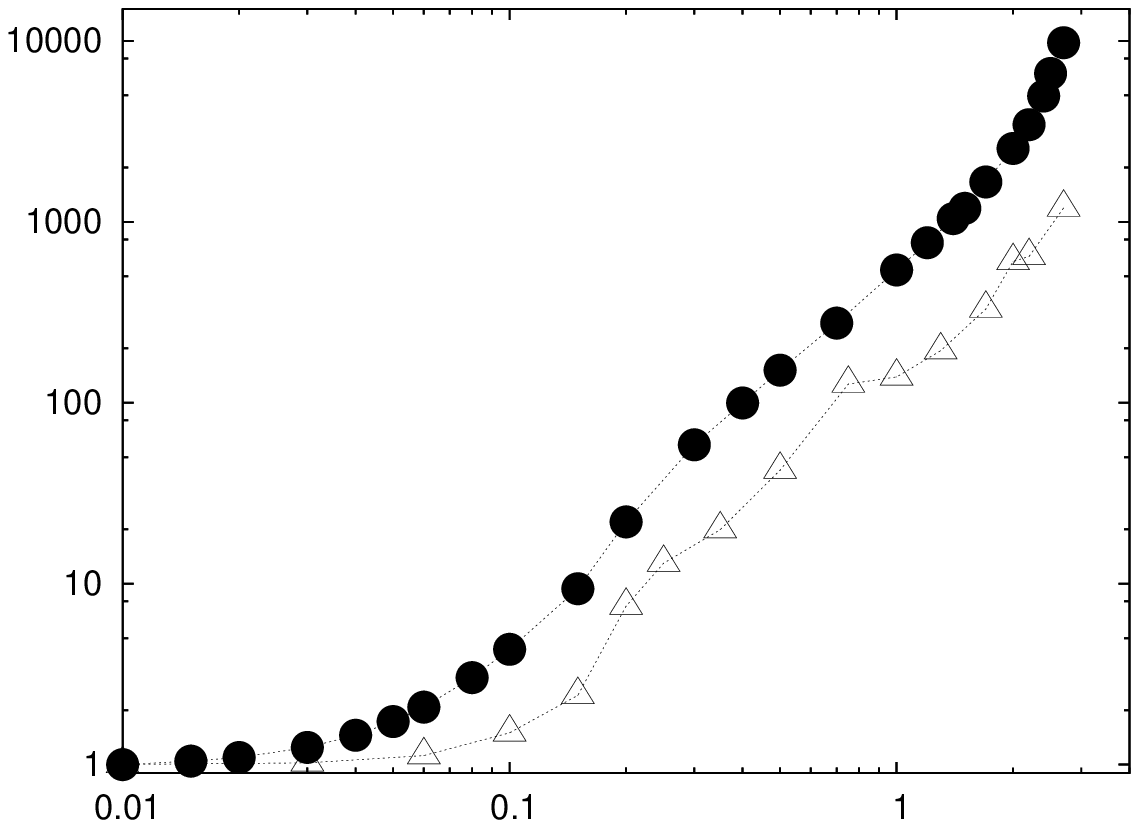} \put(-222,93){a)}
  \put(-248,38){ \rotatebox{90}{$v_w ~~~~~[\lambda_{\scriptscriptstyle
        0} / t_{\scriptscriptstyle 0}]$}}
  \put(-210,-5){$ h_{pd}/10^{\gamma_2}$}
  \put(-180,16){\rotatebox{10}{\scriptsize $0.1$}}
  \put(-205,53){\rotatebox{59}{\scriptsize $1.0$}}
  \put(-200,33){\rotatebox{34 }{\scriptsize $\gamma_2\!\!=\!\!2.0$}}
  \put(-125,90){b)} \put(-75,80){$\frac{ \mu_{\scriptscriptstyle
        I}(\gamma_2 )} {\mu_{\scriptscriptstyle I}(\gamma_2\!\!=\!\!0)} $}
  \put(-57,73){\vector(0,-1){15}} \put(-60,20){$\frac{ \langle |
      \mathcal{T}_y (\gamma_2 ) | \rangle} {\langle | \mathcal{T}_y
      (\gamma_2\!\!=\!\!0) | \rangle} $} \put(-35,35){\vector(0,1){15}}
  \put(-75,-2){$\gamma_2$}
  \caption{a) Domain wall drift velocity as a function of $h_{pd}$ for
    varying spin-orbit couplings, $\protect \gamma_2$. Note that the
    scale of the horizontal axis is different for each spin-orbit
    coupling, done to collect all graphs into a single plot. b)
    Current-driven domain wall mobility, $ \protect
    \mu_{\scriptscriptstyle I}$, and spatial total of the absolute
    values of the out-of-plane torque, $\langle | \mathcal{T}_y |
    \rangle$, as functions of spin-orbit coupling. For both plots,
    $\protect \lambda_w / \protect \lambda_{\scriptscriptstyle 0}
    \!\!=\!\! 2$.  Lines are guides for the eye.}
\label{DWVel}
\end{figure}
%
Fig.\ref{DWVel}a) shows the domain wall drift velocity as a function
of $ h_{pd}$ for varying spin-orbit coupling strength. We see that
$v_w$ is proportional to $h_{pd}$, \textit{i.e.} the domain wall drift
velocity is proportional to the current density, consistent with
experimental observations~\cite{Yamanouchi:prl06}.  In Fig.\ref
{DWVel}b), we show the current-driven domain wall mobility as a
function of spin-orbit coupling. We see that the domain wall mobility
increases four orders of magnitude when the hole spin-orbit coupling
increases from $\gamma_2 \!\!=\!\! 0$ to a typical value of $\gamma_2
\!\!=\!\! 2.7$.  In Fig.\ref{DWVel}b), we also show the spatial total
out-of-plane torque as a function of the spin-orbit coupling. We see
that $ \langle | \mathcal{T}_y (\gamma_2 ) | \rangle / \langle |
\mathcal{T}_y (\gamma_2\!\!=\!\!0) | \rangle $ increases three orders
of magnitude when $ \gamma_2$ increases from 0 to 2.7.  Increasing
spin-orbit coupling increases the out-of-plane torque and thereby the
mobility.

The effects of impurities~\cite{Timm:jpcm03} have been disregarded
here.  However, we believe that the present study still captures the
essential physics. For example, we find that the intrinsic domain wall
resistance persists in the diffusive transport
regime~\cite{Shchelushkin:prb06}. Therefore, we expect that the
spin-orbit induced effects presented in this paper are important also
for the current-driven domain wall dynamics in the diffusive transport
regime when the mean free path is not much smaller than $ \lambda_w$.
The expected impurity induced reduction of the domain wall drift
velocity is consistent with recent experimental findings~\cite
{Yamanouchi:Nature04,Yamanouchi:prl06,Gould:cm0602135}. Experiments
find a two to three orders of magnitude enhancement of the drift
velocity, which is smaller than our computed three to four order of
magnitude enhancement in the ballistic regime.

In summary, the intrinsic spin-orbit coupling in (Ga,Mn)As increases
the out-of-plane torque on the domain wall and thereby the domain wall
current-mobility by three to four orders of magnitude when the
spin-orbit coupling increases from $\gamma_2 \!\!=\!\! 0$ to an
experimentally attainable value of $\gamma_2 \!\!=\!\! 2.7$.

We thank G.\ E. W.\ Bauer, J.\ Hove, K. Olaussen, G.\ Tatara, and Y.\
Tserkovnyak for stimulating discussions. This work has been supported
by the Research Council of Norway through grants no.\ 167498/V30,
162742/V00, 1585181/143, and 1585471/431.



\end{document}